\documentclass[mr,manuscript]{copernicus}

\usepackage{mathtools}

\begin{document}
\nolinenumbers

\title{Concurrent J-Evolving Refocusing Pulses}

\author[1,2,3]{Sebastian Ehni}
\author[1,4,5]{Martin R. M. Koos}
\author[1,2,3]{Tony Reinsperger}
\author[1,2]{Jens D. Haller}
\author[1,6]{David L. Goodwin}
\author[1,2]{Burkhard Luy}

\affil[1]{Institute for Biological Interfaces 4 -- Magnetic Resonance, Karlsruhe Institute of Technology (KIT), Karlsruhe, Germany}
\affil[2]{Institute of Organic Chemistry, Karlsruhe Institute of Technology (KIT), Karlsruhe, Germany}
\affil[3]{Bruker Biospin GmbH: Ettlingen 76275, Germany; F\"{a}llanden 8117, Switzerland}
\affil[4]{Pfizer Inc.: Groton, Connecticut 06340, USA}
\affil[5]{Department of Chemistry, Carnegie Mellon University, Pittsburgh, USA}
\affil[6]{Chemistry Research Laboratory, University of Oxford, Mansfield Road, Oxford, OX1 3TA, UK}

\correspondence{David L. Goodwin (david.goodwin@partner.kit.edu) and Burkhard Luy (burkhard.luy@kit.edu)}

\runningtitle{J-Evolution Pulses}

\runningauthor{Ehni, Koos, Reinsperger, Haller, Goodwin, Luy}

\received{\today}
\pubdiscuss{} 
\revised{}
\accepted{}
\published{}

\firstpage{1}

\maketitle

\begin{abstract}
Conventional refocusing pulses are optimised for a single spin without considering any type of coupling. However, despite the fact that most couplings will result in undesired distortions, refocusing in delay-pulse-delay-type sequences with desired heteronuclear coherence transfer might be enhanced considerably by including coupling evolution into pulse design. We provide a proof of principle study for a Hydrogen-Carbon refocusing pulse sandwich with inherent J-evolution following the previously reported ICEBERG-principle with improved performance in terms of refocusing performance and/or overall effective coherence transfer time. Pulses are optimised using optimal control theory with a newly derived quality factor and z-controls as an efficient tool to speed up calculations. Pulses are characterised in detail and compared to conventional concurrent refocusing pulses, clearly showing an improvement for the J-evolving pulse sandwich. As a side-product, also efficient J-compensated refocusing pulse sandwiches -- termed BUBU pulses following the nomenclature of previous J-compensated BUBI and BEBE(tr) pulse sandwiches -- have been optimised. 
\end{abstract}

\introduction

The emergence of cryogenically cooled probe-heads, and the historic trend of increasing magnetic field strength in \textsc{nmr} (nuclear magnetic resonance) spectroscopy, present new requirements for the design and implementation of effective \textsc{nmr} experiments. Traditional non-selective hard pulses barely cover the required bandwidths of typical heteronuclei. For example for \(^{13}\)C spectroscopy, the uniform coverage of approximately \(37.5~\unit{kHz}\) on a routine \(600~\unit{MHz}\) \textsc{nmr} spectrometer is required, which is already a challenge. The formulation of composite pulses \citep{Levitt:1982} increased the pulse sequence bandwidth, covering a large resonant frequency range \citep{Shaka:1983,Warren:1984,Tycko:1985,Levitt:1986,Freeman:1988}. Indirectly, this success led to the development of numerical optimisation to engineer broadband composite pulses \citep{Lurie:1985,Conolly:1986,Shaka:1987,Emsley:1990,Ewing:1990,Garwood:1991}. A similar development took place regarding radio-frequency pulses compensated for field inhomogeneities: when used with cryogenic probeheads with very high sensitivity to salt concentrations and typically slightly increased B\(_1\)-inhomogeneities, optimised pulses require an increased range of effective pulse strengths, typically covering variations in \(B_1 \ge \pm 20\%\). 

Numerically optimised pulse shapes \citep{Glaser:2015} increased the effectiveness of \textsc{nmr} spectroscopy with the same objective as composite pulses: tolerance to a range of pulse strengths. An explosion of pulse engineering began with the possibility to optimise pulse shapes containing thousands of independent variables, \emph{e.g.} pulse amplitudes and phases \citep{Khaneja:2005}. Optimal control theory has proved an indispensable tool for the optimisation of shaped pulses. At the present time, pulses are available for a variety of nuclei and bandwidths such as \(^1\)H and \(^{13}\)C, but also for more specialised applications \emph{e.g.} slice-selective and low-power \textsc{mri} (magnetic resonance imaging) pulses \citep{Janich:2011,Vinding:2012,Vinding:2017,VanReeth:2017}, culminating in utmost complex pattern pulses \citep{Kobzar:2005}, and hardware distorted microwave pulses for the use in \textsc{epr} (electron paramagnetic resonance) spectroscopy \citep{Spindler:2012,Doll:2013,Kaufmann:2013,Goodwin:2018}. Optimal pulse engineering is leading to the question whether optimal pulses can approach the physical limit for demanding pulse robustness. Systematic studies on the optimisation of shaped pulses \citep{Kobzar:2004,Kobzar:2008,Kobzar:2012} and on quantum evolution with known physical limits \citep{Reiss:2002,Khaneja:2002a,Khaneja:2002b,Reiss:2003,Khaneja:2003a,Khaneja:2003b,Stefanatos:2004} lead to time optimal curves that are a versatile tool to find estimates for physical limits in spin dynamics \citep{Kobzar_thesis}.

As the limit of what is physically possible is neared, and control problems become computationally arduous, the development of optimal control methods is essential. There are a number of different approaches to optimal control: time optimal control \citep{Khaneja:2001}; problems solved numerically with a piecewise-constant pulse approximation \citep{Skinner:2003,Skinner:2004,Skinner:2005} with linear \citep{Khaneja:2005}, super-linear \citep{deFouquieres:2011}, or quadratic convergence \citep{Goodwin:2016}; annealing a spin system to a desired effective Hamiltonian \citep{Tosner:2006}; utilisation of cooperative multi-pulse control \citep{Braun:2014}; optimal control using analytic controls \citep{Machnes:2018}. Specifically, the optimal control method of \textsc{grape} (gradient ascent pulse engineering) \citep{Khaneja:2005} is used for diverse applications \citep{Hohenester:2005,Palao:2008,Ndong:2010,Spindler:2012,Dolde:2014,Saywell:2018}, and can be used to construct \emph{universal rotation} (UR) pulses, being unitary propagators of the system \citep{Kobzar:2012,Skinner:2012,Dolde:2014,Lingel:2020}.

The design of time optimal experiments \citep{Khaneja:2001} for time efficient coherence transfer elements \citep{Kobzar:2004,Ehni:2012,Kobzar:2012} is equally important because most of the experiment time should consist of inter-pulse delays, used for coherence transfer or chemical shift evolution, rather than radio-frequency pulsing. The simple conclusion is to ensure that total pulsing duration is as short as possible. Ideally, the total sequence duration should not be increased by the duration of the shaped pulse, \(T\). This would be equivalent to the overall optimisation of the \textsc{nmr} experiment with respect to all relevant interactions from the underlying Hamiltonian \(\mathbf{H}\), \emph{e.g.} pulse strength and offset deviations of all nuclei involved in the spin system, including spin-spin coupling.

An indirect way to obtain the \emph{sequence of pulses and delays} has been proposed with the \textsc{iceberg}-principle \citep{Gershenzon:2008}: optimised pulse shapes followed directly by a time period of free evolution. Such pulses are named \textsc{iceberg} pulses as only a tiny fraction of the pulse contributes to the overall pulse length (\emph{in view, above the sea level}), whilst the main part of the pulse can be considered part of the required delay time (\emph{not in view, below the sea level}). These pulses have been generalised to arbitrary flip-angle \textsc{radfa} pulses \citep{Koos:2015,Koos:2017}, and used e.g. in Ramsey-type cooperative experiments \citep{Braun:2014}. The novel feature of \textsc{iceberg} and \textsc{radfa} pulses, in the context of optimal control, is that each spin in the offset ensemble has a uniquely defined terminal phase, and the ensemble produces a linear phase dispersion. In addition, UR pulses producing quadratic phase dispersion (\textsc{sordor}) have recently been proposed by \citet{Goodwin:2020}.

This phase is an effective \(z\)-rotation, and assuming radio-frequency controls dominate the Hamiltonian \(\mathbf{H}\), the phase evolution can be mimicked by controls. Considering this relationship between frequency offset and controls, the \textsc{iceberg} pulses can be described as the simplest in a class of \emph{drift Hamiltonian mimicking} pulses. Furthermore, setting an optimal control problem to incorporate both the shape of pulses and coherence transfer elements may reveal a path to time optimal experiments \emph{e.g.} in heteronuclear spectroscopy, the effect of two concurrent UR-\(180\degree\) pulses set between delays is to refocus chemical shift evolution during coherence transfer (\textsc{cob-hsqc}, \citep{Ehni:2012}) whilst retaining maximal J-evolution. This class of pulse will be termed J-evolution, or J\(_\text{ev}\), pulses in the following.

\section{Theory}

\subsection{J-Evolution and Fidelity Function}

A system irradiated in a pulsed experiment can be described as a bilinear control system,
\begin{equation}
\mathbf{H}(t) = \mathbf{H}_0+\sum\limits_{\ell}^{L}c_\ell(t)\mathbf{H}_\ell
\end{equation}
where \(\mathbf{H}_0\) describes the uncontrollable parts of the system, termed the \emph{drift Hamiltonian}. The controls of the system, \(\mathbf{H}_\ell\), are associated with real-valued, time-dependent, control irradiation amplitudes, \(c_\ell(t)\).

A system of \(K\) coupled spins in isotropic solution can be described by each of their resonant frequency offsets, \(\omega_{k}\), and a scalar coupling term, \(J_{k,k^{\prime}}\), linking each pair of spins \(k\) and \(k^{\prime}\). When considering only heteronuclear systems, the weak coupling approximation can be used, which gives the drift Hamiltonian as
\begin{equation}
\mathbf{H}_0 = \sum\limits_{k=1}^{K} \omega_{k} \mathbf{H}_\mathrm{z}^{(k)} + \sum\limits_{k^{\prime}>k}^{K} 2\pi J_{k,k^{\prime}} \Big(\mathbf{H}_\mathrm{z}^{(k)}\mathbf{H}_\mathrm{z}^{(k^{\prime})}\Big)
\label{EQ_BilinearHamiltonian}
\end{equation} 
where Pauli multi-spin operators are denoted \(\mathbf{H}_{\mathrm{x},\mathrm{y},\mathrm{z}}\).

The \textsc{grape} method of optimal control \citep{Khaneja:2005} proceeds to describe the irradiation as piecewise constant over a small time interval \(\Delta t\). This approximation allows the numerical solution of Eq.~(\ref{EQ_BilinearHamiltonian}) with calculation of the time propagator at each time interval, \(\mathbf{P}_n\). Consequentially, the effect of Hamiltonian dynamics from a time \(t_0\) to a particular time \(t_n\) is described by the time-ordered effective propagator, \(\mathbf{U}_n\), starting from the initial propagator of the system  \(\mathbf{P}_0\), usually the identity matrix, and \(\mathbf{V}_n\) are the effective propagators of the adjoint control problem, propagated with reverse time-order
\begin{align}
& \mathbf{P}_n^{}=\exp{\big\{-i\mathbf{H}_n^{}\Delta t\big\}}, &
  \begin{array}{l}
   \mathbf{U}_n=\mathbf{P}_n\dots\mathbf{P}_1\mathbf{P}_0\\
   \mathbf{V}_n=\mathbf{P}_{n+1}^{\dagger}\dots\mathbf{P}_N^{\dagger}\mathbf{R}.
  \end{array} & \label{EQ_EffectivePropagators}
\end{align}

J-coupling evolution can be included in the description of frequency offset and/or \(B_1\) compensated pulses \citep{Ehni:2012,Kobzar:2012,Ehni:2013,Ehni:2014}, mapping both into J-robustness. A measure to numerically optimise such pulses is the distance between a desired effective propagator, \(\mathbf{R}\), and the implemented effective propagator, \(\mathbf{U}_N\), using \(\big\|\mathbf{R}-\mathbf{U}_N\big\|\) \citep{Khaneja:2005,Tosner:2006}, which can be expressed as the maximisation of their scalar product (Hilbert-Schmidt inner product)
\begin{equation}
 \max_{c_\ell}\Phi=\max_{c_\ell}\Big[\mathrm{Re}\big\langle \mathbf{R}\big|\mathbf{U}_N\big\rangle\Big]\label{EQ_maxFidelity}.
\end{equation}

A broadband pulse sequence, additionally robust with respect to scaling of the radio-frequency amplitude, is defined by the average over a set of \(n_\mathrm{off}\) equally spaced offsets \(\omega_\mathrm{o}\) and \(n_\mathrm{rf}\) equally spaced scaled maximum radio-frequency amplitudes \(\omega_\mathrm{rf}\) and \(n_J\) equally spaced J-couplings in the desired range of offset, radio-frequency scaling and J-coupling \citep{Khaneja:2005}.

In this way, two concurrent J-compensated \(180\degree\) broadband UR pulses (abbreviated as \textsc{bubu}-pulse -- a \emph{Broadband Universal, Broadband Universal} sandwich) can be found with optimal control with the desired target propagator
\begin{equation}
\mathbf{R}_{\pi}=\exp{\Big\{-i\pi\big(\mathbf{H}_\mathrm{x}^{(1)}+\mathbf{H}_\mathrm{x}^{(2)}\big)\Big\}}
\label{Eq_PropTarg}
\end{equation}

\begin{figure}
\centering{\includegraphics{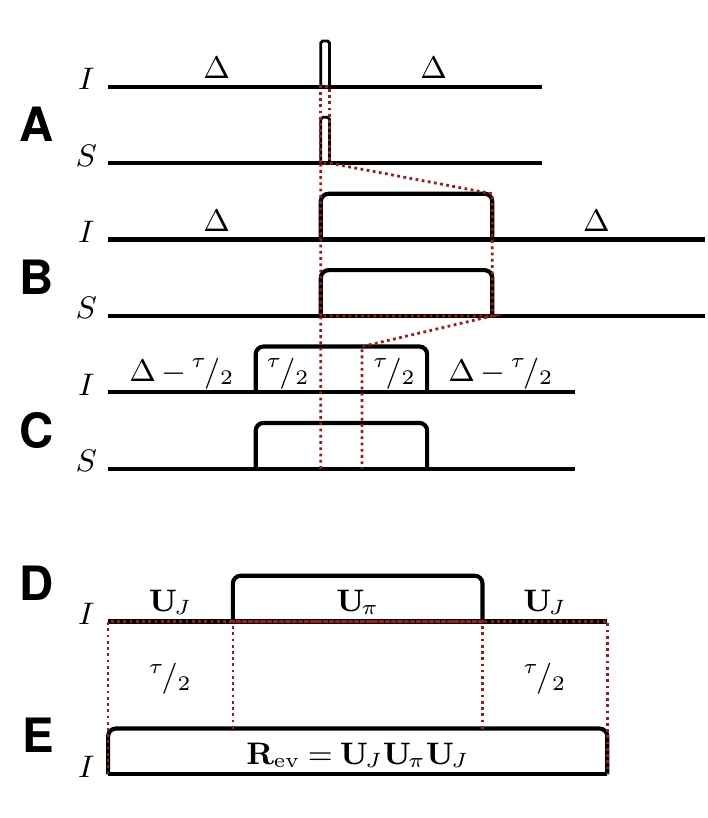}}
\caption{Spin-echo pulse sequences for the refocusing of chemical shift during heteronuclear coupling evolution with respect to coupled nuclei \(H^{(k)}\) and \(H^{(k^{\prime})}\). With (A) concurrent hard pulses, (B) concurrent shaped pulses that increase the total sequence duration to \(= 2\Delta + T\), (C) concurrent shaped pulses with total sequence duration decreased to \(= 2\Delta + T - \tau\) that can be imagined as a delay \(\tau\) that is moved inside the shaped pulse to result in a J\(_\text{ev}\) pulse. (D) J\(_\text{ev}\) pulses are comparable to J\(_\text{ev}\)-building blocks that are composed from \textsc{bubu} refocusing pulses and delays \(\tau\). A J\(_\text{ev}\) pulse (E) is made up of the product of unitary rotations. \(\mathbf{R}_J\) is the unitary propagator that results from a free precession period \(\tau/2\). This is combined with the effect of a refocusing pulse \(\mathbf{R}_{\pi}\) to form \(\mathbf{R}_\mathrm{ev}\) that is the final propagator of a J\(_\text{ev}\) pulse.\label{FIG_Pulse_sequences}}
\end{figure}

Optimisations using \(\Phi\) yield robust analogues (Fig.~\ref{FIG_Pulse_sequences}(B)) of concurrent hard \(180\degree\) pulses. These refocusing pulses are used in the centre of delays whereupon the total sequence duration is increased by the length of the shaped pulse \(T\) (as indicated by the dotted lines in Fig.~\ref{FIG_Pulse_sequences}).

A further assertion is made: shorter sequences can be obtained when the delays are partly moved into the shaped pulse (Fig.~\ref{FIG_Pulse_sequences}(C)). A fraction of time, indicated in Fig.~\ref{FIG_Pulse_sequences}(C), is defined by
\begin{align}
& \tau  = q T, & q\in [0 ,1]. &\label{EQ_constant_q}
\end{align}
In this case, the total sequence duration would be decreased by \(\tau\), and is equivalent to the idea that J-coupling evolution is mediated through the pulse. This concept is realised by a new fidelity function \(\Phi_\mathrm{ev}\). Accordingly, pulses obtained by \(\Phi_\mathrm{ev}\) are termed J-evolution, or simply J\(_\text{ev}\), pulses.
\begin{equation}
\Phi_\mathrm{ev}=\mathrm{Re}\big\langle \mathbf{R}_\mathrm{ev}\big| \mathbf{U}_\mathrm{eff}\big\rangle\label{EQ_fidelity_ev}
\end{equation}
with
\begin{equation}
\mathbf{R}_\mathrm{ev}=\mathbf{R}_J\mathbf{R}_{\pi}\mathbf{R}_J\label{Eq_prop_ev}
\end{equation}
and
\begin{equation}
\mathbf{R}_J=\exp{\bigg\{-i2\pi J\mathbf{H}_\mathrm{z}^{(1)}\mathbf{H}_\mathrm{z}^{(2)}\frac{qT}{2}\bigg\}}\label{Eq_prop_J}
\end{equation}

The target propagator \(\mathbf{R}_\mathrm{ev}\) (Fig.~\ref{FIG_Pulse_sequences}(E)) can be written as the product of \(\mathbf{R}_J\) and \(\mathbf{R}_{\pi}\). It can be imagined as a pulse that mediates the effect of a free precession period \(\tau/2\) followed by an ideal \(180\degree\) concurrent refocusing pulse followed by one more \(\tau/2\) precession period as depicted in Fig.~\ref{FIG_Pulse_sequences}(D). The optimisation needs to accommodate that in a shaped pulse.

\subsection{Calculation of Gradients}

The gradient-following \textsc{grape} method of optimal control requires directional propagator derivatives at each time increment, \(D_{\mathbf{\sigma}}(\mathbf{P}_n^{})\), in the direction of each control operator. For phase modulated pulses, the gradient vector is constructed from the elements \citep{Skinner:2006}
\begin{equation}
 \frac{\partial \Phi}{\partial \varphi_{n}^{}}=x_n^{}\big\langle\mathbf{V}_n\big|D_{\mathbf{\sigma}_\mathrm{y}}(\mathbf{P}_n^{})\mathbf{U}_n\big\rangle - y_n^{}\big\langle\mathbf{V}_n\big|D_{\mathbf{\sigma}_\mathrm{x}}(\mathbf{P}_n^{})\mathbf{U}_n\big\rangle
 \label{EQ_DirectionalDerivatives}
\end{equation}
The gradient needed for the optimisation is calculated as the derivative of the fidelity function \(\Phi_\mathrm{ev}\) with respect to the \(x\)- and \(y\)-controls.
\begin{equation}
\frac{\partial \Phi}{\partial c_k(t_n)}=\frac{\partial \Phi}{\partial c_{n,k}}=\mathrm{Re}\Bigg\langle \mathbf{P}_{n+1}^{\dagger}\cdots \mathbf{P}_N^{\dagger} \mathbf{U}_\text{eff}^{} \Bigg| \frac{\partial \mathbf{P}_n}{\partial c_{n,k}}\cdots \mathbf{P}_1^{}\Bigg\rangle
\end{equation}
where the challenging task is to calculate the derivative of \(\mathbf{P}\), which is given in a first order approximation by \citep{Khaneja:2005}
\begin{equation}
\frac{\partial \Phi}{\partial c_{n,k}}=-\mathrm{Re}\Big\langle \mathbf{P}_{n+1}\Big| i\Delta t \mathbf{H}_k^{}\mathbf{U}_n^{}\Big\rangle
\end{equation}

For the present study, however, the exact derivative of \(\mathbf{P}\) is used that can be obtained, apart from other methods \citep{Goodwin:2015}, by an eigenbasis transformation into the basis of the time independent Hamiltonians \(\mathbf{H}\) \citep{Levante:1996}.

In the basis of \(\mathbf{H}\) the matrix exponential \(\mathbf{P}\) and its derivatives, with respect to the controls \(c_k\), collapse to a scalar exponential and its ordinary derivatives. Using the product rule, the derivatives of the propagator, with respect to the controls, are obtained by
\begin{align}
\frac{\partial \mathbf{U}_n}{\partial c_{n,k}} & = \frac{\partial}{\partial c_{n,k}}\mathrm{e}^{-i\mathbf{H}_nt} = \frac{\partial}{\partial c_{n,k}}\mathbf{Q}\mathrm{e}^{-i\mathbf{\Lambda}_nt}\mathbf{Q}_{}^{\dagger}\nonumber\\
&=\frac{\partial \mathbf{Q}}{\partial c_{n,k}}\mathrm{e}^{-i\mathbf{\Lambda}_nt}\mathbf{Q}_{}^{\dagger} + \mathbf{Q}\frac{\partial \mathrm{e}^{-i\mathbf{\Lambda}_nt}}{\partial c_{n,k}}\mathbf{Q}_{}^{\dagger} + \mathbf{Q}\mathrm{e}^{-i\mathbf{\Lambda}_nt}\frac{\partial \mathbf{Q}_{}^{\dagger}}{\partial c_{n,k}}\nonumber\\
&=\mathbf{Q}\frac{\partial \mathrm{e}^{-i\mathbf{\Lambda}_nt}}{\partial c_{n,k}}\mathbf{Q}_{}^{\dagger}
\end{align}
where \(\mathbf{\Lambda}_n\) is a diagonal matrix containing the eigenvalues of \(\mathbf{H}_n\) and \(\mathbf{Q}\) is the matrix
with columns of corresponding eigenvectors. The only non-zero derivative is that of the exponential function.

\subsection{Optimisation Parameters}\label{optpara}

The initial pulse \(c_0\) is chosen as random controls. Conjugate gradients are used for update, and controls are truncated once they exceed \(\omega_\mathrm{rf} = 20~\unit{kHz}\) for \(^1\)H and \(\omega_\mathrm{rf} = 10~\unit{kHz}\) for \(^{13}\)C.

The optimisation parameters are chosen to accommodate the common bandwidths and \(B_1\) field inhomogeneities for \(^1\)H and \(^{13}\)C spin systems, respectively by setting \(\Delta \omega_{k} =7~\unit{kHz}\) (8), \(\Delta \omega_{k^{\prime}} = 37.5~\unit{kHz}\) (42), \(B_1\)-compensations according to \(\theta_{k} = \pm 20\%\) (3), \(\theta_{k^{\prime}} = \pm 5\%\) (2), and a J-coupling up to \(250~\unit{Hz}\) (1). The number of explicit and equidistant evaluations of the fidelity and gradient functions are given in parentheses. Optimisation and evaluation parameters are identical except that the number of full time propagators in the respective dimension of the evaluation is somewhat higher to avoid any dependence of the fidelity function on the number of points used for evaluation.

\subsection{z-Controls}

A practical limitation of the optimisation is the large number of \(\omega_1\), \(\omega_2\), corresponding B\(_1\)-inhomogeneities, J-coupling, and timesteps for a realistic pulse shape digitisation of \(0.5~\unit{\mu s}\) which easily results in several million propagation calculations per optimisation step. Reducing digitisation towards longer timesteps is not acceptable for the desired bandwidths: 
assuming a maximum carbon bandwidth of \(37.5~\unit{kHz}\) there is a phase evolution of up to approximately 67.5\degree within an increment duration of \(\Delta t = 10~\unit{\mu s}\). This unnecessarily brings a strong limitation to the available radio-frequency controls that can limit the fidelity obtained by a \(\Phi_\mathrm{ev}\)-optimisation.

To avoid that, an additional set of controls, namely \(z\)-controls can be introduced. \(z\)-controls are calculated in accordance to the formulas derived for \(x\), \(y\)-controls. \(z\)-controls cause \(z\)-rotations thereby allowing the application of frequencies with lower digitization than with conventional \(x\),\(y\)-controls. They can be imagined as the effect that occurs on changing the transmitter offset to a new value. Since any \(z\)-rotation can be mediated also by a phase sweep of \(x\), \(y\)-controls, it is possible to transform \(x\), \(y\), \(z\)-pulses into \(x\), \(y\)-pulses of higher digitisation that can be realised by the available hardware. By this, the difference between high resolution \(x\), \(y\) and lower resolution \(x\), \(y\), \(z\) pulses is reduced to a distinct incrementation of radio-frequency offset changes, while the whole range of possible radio-frequency offsets is not reduced. As such, after a pulse is obtained the \(z\)-controls are rendered with a resolution of \(0.5~\unit{\mu s}\) onto the \(\Delta t = 10~\unit{\mu s}\) delays to obtain a pulse that can be used on the spectrometer. Nonetheless, simulations given in the following directly use the obtained \(z\)-controls for simplicity.

\section{Results and Discussion}

\subsection{Pulses Obtained}

\begin{figure}
\centering{\includegraphics{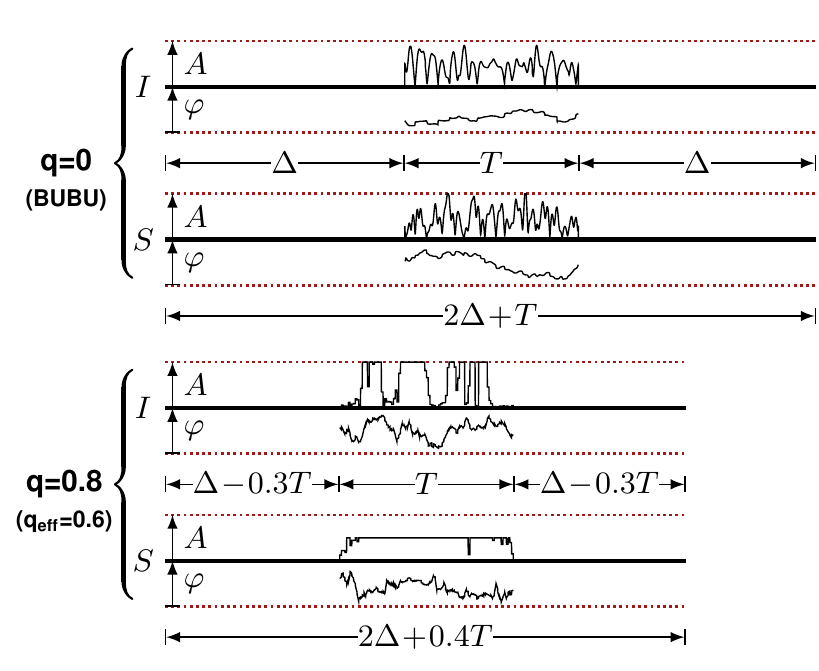}}
\caption{Amplitude, \(A\), and phase, \(\varphi\), of example J\(_\text{ev}\) pulse sandwich with \(T=1\)~ms duration, optimised for an effective \(180\degree\) refocusing on both \(I\) and \(S\) nuclei and coupling evolution according to the time fractions \(q=0\) (\textsc{bubu}) and \(q=0.8\) (equivalent to \(q_\mathrm{eff}\approx0.6\) in Eq.~(\ref{Eq_qeff}) of Sect.~\ref{sect_qeff}). Diagrams are in the same style as Fig.~\ref{FIG_Pulse_sequences}(C). The scale of the pulse amplitude is shown with the dotted lines at \(20~\unit{kHz}\), and the pulse phases are arbitrarily unwrapped with the scale between \(0\) and a maximum of \(5\times 2\pi\) turns.\label{FIG_pulses}}
\end{figure}

Using the optimisation parameters defined in Sect.~\ref{optpara}, a total of five J\(_\text{ev}\) pulse sandwiches of 1~ms duration with \(q \in\{0.2, 0.4, 0.6, 0.8, 0.999\}\) for \(^1\)H,\(^{13}\)C-experiments were optimised. As an example, the \textsc{bubu} pulse sandwich (with \(q=0\)) and \(q=0.8\) (equivalent to \(q_\mathrm{eff}\approx0.6\) in Eq.~(\ref{Eq_qeff}) of Sect.~\ref{sect_qeff}) are shown in Fig.~\ref{FIG_pulses}. The other pulses look qualitatively similar.

As the pulses are designed to accommodate both J-coupling evolution and a \(180\degree\) rotation, a proper evaluation of the  J\(_\text{ev}\) pulse sandwiches involve different comparisons as will be shown in the following sections. 

\subsection{Detailed Evaluation of Example Pulses}

Conventionally, pulses are evaluated according to their offset dependence of transfer properties to indicate their overall performance. For the pulse sandwiches optimised here, an overall offset- and J-dependent target propagator \(\mathbf{R}_\mathrm{ev}(\theta_1,\theta_2,\omega_1,\omega_2,J)\) is evaluated in the quality factor \(\Phi_\mathrm{ev}\). In a first step,  \(\Phi_\mathrm{ev}\) is calculated for different offsets and B\(_1\) values: the refocusing and J-evolution is essentially fulfilled over the desired \(\omega_1\) offset-range and \(\theta_1\) inhomogeneity-range, which is also the case for the overall range of \(\omega_2\) offsets, \(\theta_2\) inhomogeneities, and J-couplings.  

The direct effect from optimised pulses over the pulse duration is not useful and it should be expected that trajectories show an apparently disordered, even noisy, trajectory which starts from a defined state and finishes in a defined state, for each offset in the optimised range. This randomised feature is very different to pulse sandwiches designed by hand, as e.g. the case for selective \textsc{reburp} pulses within selective \textsc{inept} steps \citep{Lescop:2010,Haller:2019,Bodor:2020}. Nevertheless, despite the disordered behaviour, the resulting performance over the offset range is good.

\begin{figure*}
\centering{\includegraphics{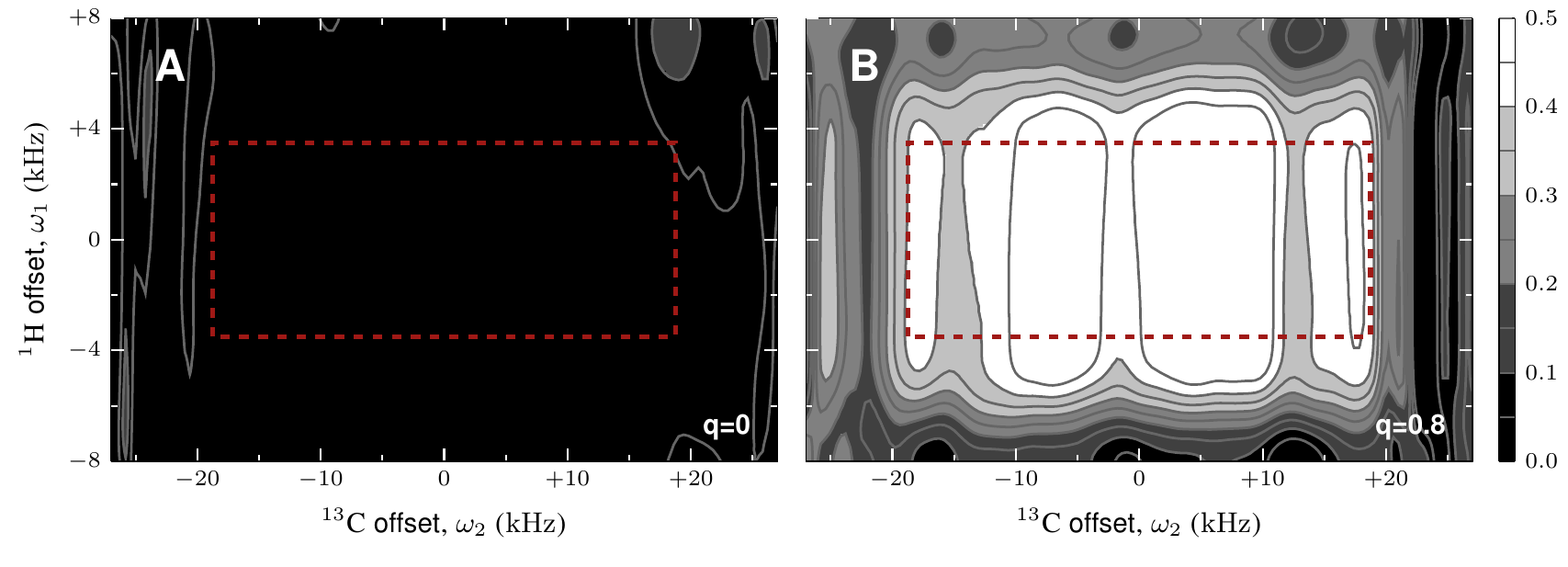}}
\caption{First order toggling frame based analysis of the scaling of heteronuclear coupling Hamiltonian with respect to \(^1\)H offsets, \(\omega_1\), and \(^{13}\)C offsets, \(\omega_2\). Scaling factors for optimised J\(_\text{ev}\) pulse sandwiches with (A) \(q=0\) (\textsc{bubu}) and (B) \(q=0.8\) (equivalent to \(q_\mathrm{eff}\approx0.6\) in Eq.~(\ref{Eq_qeff}) of Sect.~\ref{sect_qeff}), and are shown in greyscale using the scale on the right side. The dashed box indicates the region optimised for J\(_\text{ev}\) pulses. A clear island of J-evolution emerges at the higher \(q\) value.\label{FIG_Toggling_frame}}
\end{figure*}

Following the analysis of selective refocusing pulses in heteronuclear correlation experiments \citep{Haller:2019}, also an (\(\omega_1,\omega_2\)) offset dependent profile of effective J-evolution has been calculated using a toggling-frame approach with explicit inclusion of offset-frequencies. Essentially the scaling of the longitudinal heteronuclear coupling Hamiltonian \(\mathbf{H}_J=2 \pi J \ \mathbf{H}_z^{(1)} \cdot \mathbf{H}_z^{(2)}\) is calculated, which is directly related to the fraction \(q\) of the target propagator. Unfortunately, the relatively complicated computation is beyond the scope of this publication and will be explained in detail in a different article (manuscript in preparation). As can be seen from the results shown in Fig.~\ref{FIG_Toggling_frame}, a relatively homogeneous coupling evolution is achieved which, however, does not fully reach the fraction \(q\) specified in the target propagator, indicating the physical limits of J-evolution. Fig.~\ref{FIG_Toggling_frame}(A) shows the coupling evolution for a low \(q\) value, with no pattern and very little coupling evolution over the optimised range, which should be expected. This is in stark contrast to Fig.~\ref{FIG_Toggling_frame}(B), for a high \(q\) value, where a clear island of coupling-evolution emerges from an ocean of no coupling-evolution. A comparison of the different pulse sandwiches will be shown in the following sections using different overall fidelities to better quantify the various aspects of pulse performance.

\subsection{Overall Performance of Pulse Sandwiches}\label{sect_qeff}

\begin{figure*}
\centering{\includegraphics{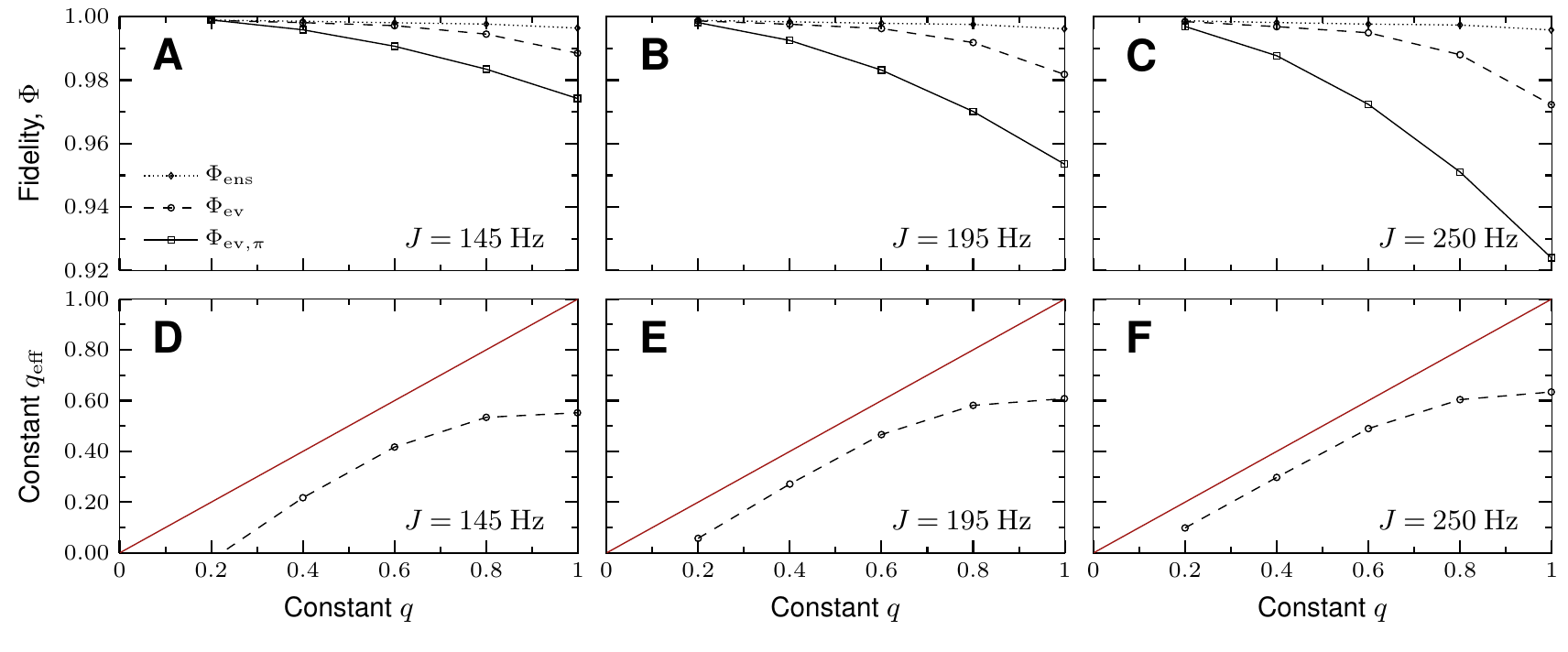}}
\caption{Transfer efficiencies for concurrent \(180\degree\) refocusing pulses, calculated for \(J \in \{ 145, 195, 250 \}~\unit{Hz}\) and averaged over offsets and \(B_1\) deviations. Five J\(_\text{ev}\) pulses are evaluated as a function of the J-evolution constant \(q\) which demands a ratio of J-coupling that is based on the pulse length \(T = 1~\unit{ms}\). (A-C) Upper limit of transfer efficiency that can be reached by conventional, concurrent refocusing pulses is calculated by \(\Phi_\mathrm{ev}\) (solid lines). (A-C) J\(_\text{ev}\) pulses acquire heteronuclear J-coupling and exceed that limit (dashed lines). (A-C) The \(180\degree\) performance of J\(_\text{ev}\) pulses is obtained by the fidelity function \(\Phi_\mathrm{ens}\) (dotted lines). (D-F) The J-coupling that has effectively been acquired upon a J\(_\text{ev}\) pulse is given by \(q_\mathrm{eff}\) (dashed lines). For \(q \leqslant 0.6\) the linear behaviour and the shift according to the diagonal indicates that J-evolution performance depends on \(q\).\label{FIG_Transfer_efficiencies_1}}
\end{figure*}

A comparison of pulse performance for the J\(_\text{ev}\) pulse sandwiches, obtained with \(q \in\{0.2, 0.4, 0.6, 0.8, 0.999\}\), requires an evaluation of a fidelity metric. Here, the fidelity of Eq.~(\ref{Eq_prop_ev}) is initially evaluated according to Eq.~(\ref{Eq_PropTarg}), and is shown as a function of \(q\) in Fig.~\ref{FIG_Transfer_efficiencies_1}(A-C) (dashed lines). The evaluation is done three times for different coupling constants \(J \in \{ 145, 195, 250 \}~\unit{Hz}\), representing classical aliphatic as well as aromatic and triple bond scenarios. The selectivity of an arbitrary property depends on the reciprocal of the pulse length -- on the order of \(1/1~\unit{ms} = 1000~\unit{Hz}\) for the presented J\(_\text{ev}\) pulses. Accordingly, Fig.~\ref{FIG_Transfer_efficiencies_1}(A-C) (dashed lines) show a similar behaviour because the selectivity spans the whole range of coupling constants (\(J \in \{145, 195, 250\}~\unit{Hz}\)) and approximately differ linearly as a function of the actual J-coupling constants.

Since the fidelity function \(\Phi_\mathrm{ev}\) is intended to be a measure for both \(180\degree\) pulses and J-evolution, these two properties need to be resolved, in order to judge the optimised pulses, according to their J-evolution capability. As a first step we want to evaluate how far J-evolution takes place and how it contributes to the overall performance. Therefore, an ideal effective propagator \(\mathbf{U}_{\mathrm{eff},\pi}=\mathbf{R}_\pi\) is defined by Eq.~(\ref{Eq_PropTarg}) which is evaluated according to \(\mathbf{R}_\mathrm{ev}\) in Eq.~(\ref{Eq_prop_ev}), using \(q\) of the pulse introduced in Eq.~(\ref{Eq_prop_J}), with
\begin{equation}
\Phi_{\mathrm{ev},\pi}=\mathrm{Re}\big\langle \mathbf{R}_\mathrm{ev}\big| \mathbf{U}_{\mathrm{eff},\pi}\big\rangle.\label{EQ_fidelity_evpi}
\end{equation}

The numerical values of the idealised fidelity function are presented as a plot in Fig.~\ref{FIG_Transfer_efficiencies_1}(A-C) (solid line). They assign the fidelity that would be reached upon \(\Phi_\mathrm{ev}\) by ideal concurrent \(180\degree\) pulses, i.e. infinitely hard pulses without radio-frequency variation. This class of pulse is not supposed to evolve J-coupling, which is the reason for the declining transfer efficiency with increasing \(q\) and \(J\) as illustrated in Fig.~\ref{FIG_Transfer_efficiencies_1}(A-C) (solid lines). Such ideal pulses set up a maximum fidelity that can be reached with conventional refocusing pulses. Every pulse that reports a transfer efficiency higher than this threshold must evolve J-coupling according to \(\Phi_\mathrm{ev}\). This is observed for the discussed J\(_\text{ev}\) pulses, indicating clearly that desired J-evolution during the pulse sandwich takes place. For a better quantification on the amount of overall coupling evolution, another quantity is derived by considering the threshold given by ideal \(180\degree\) pulses (Fig.~\ref{FIG_Transfer_efficiencies_1}(A-C), solid line): it is concluded that the range of values given by \(\Phi_\mathrm{ev}\) is dominated by the refocusing property. Therefore, the analysis of J-coupling properties is restricted to the range of \(1 \leqslant \Phi_{\mathrm{ev},\pi}(q)\). The difference between \(\Phi_\mathrm{ev}\) and \(\Phi_{\mathrm{ev},\pi}\) is division by \(\Phi_{\mathrm{ev},\pi}\) resulting in a fraction of the J-coupling that is acquired. This can be written as a function of \(q\) to result in a constant \(q_\mathrm{eff}\) that is actually provided by the J\(_\text{ev}\) pulse according to
\begin{equation}
q_\mathrm{eff}=q\frac{\Phi_\mathrm{ev}-\Phi_{\mathrm{ev},\pi}}{1-\Phi_{\mathrm{ev},\pi}}.\label{Eq_qeff}
\end{equation}

The values of \(q_\mathrm{eff}\) are plot as a function of \(q\) in Fig.~\ref{FIG_Transfer_efficiencies_1}(D-F), and shown explicitly in Table~\ref{tab_qeff}. This plot of \(q_\mathrm{eff}\) is nearly parallel to the diagonal for \(q \lessapprox 0.6\) indicating an increase of J-coupling capability as demanded by the coefficient \(q\). The parallel displacement of \(q_\mathrm{eff}\) to the diagonal is explained by the \(180\degree\) property of the J\(_\text{ev}\) pulse sandwich: whilst the magnetisation must be in the transverse plane to acquire J-coupling, it must also leave the transverse plane to facilitate a \(180\degree\) rotation. The shift is caused by the fraction of time within the shaped pulse that is used for the \(180\degree\) rotation. With the slope of the \(q_\mathrm{eff}\) plot parallel to the diagonal there is an indication that there are \emph{unused resources} for \(q \lessapprox 0.6\), which could be used for J-coupling evolution.

\begin{table}
\caption{The constant \(q_\mathrm{eff}\) for a number of J-coupling values calculated using Eq.~\ref{Eq_qeff} for the constant \(q\) from Eq.~\ref{EQ_constant_q}, from the two fidelity measures \(\Phi_\mathrm{ev}\) and  \(\Phi_{\mathrm{ev},\pi}\) from Eq.~\ref{EQ_fidelity_ev} and Eq.~\ref{EQ_fidelity_evpi} respectively.}\label{tab_qeff}
\vskip4mm
\begin{tabular}{l c c c r}
\tophline
\(q\) & \(J\) & \(\Phi_\mathrm{ev}\) & \(\Phi_{\mathrm{ev},\pi}\) & \(q_\mathrm{eff}\)\\
\middlehline
 0.2 & 145 & 0.9988 & 0.9990 & -0.0271 \\
 0.2 & 195 & 0.9987 & 0.9981 &  0.0576 \\
 0.2 & 250 & 0.9984 & 0.9969 &  0.0986 \\
 0.4 & 145 & 0.9981 & 0.9959 &  0.2176 \\
 0.4 & 195 & 0.9976 & 0.9925 &  0.2715 \\
 0.4 & 250 & 0.9968 & 0.9877 &  0.2976 \\
 0.6 & 145 & 0.9972 & 0.9907 &  0.4172 \\
 0.6 & 195 & 0.9962 & 0.9832 &  0.4663 \\
 0.6 & 250 & 0.9949 & 0.9724 &  0.4899 \\
 0.8 & 145 & 0.9945 & 0.9834 &  0.5342 \\
 0.8 & 195 & 0.9918 & 0.9701 &  0.5814 \\
 0.8 & 250 & 0.9880 & 0.9511 &  0.6039 \\
 0.999 & 145 & 0.9885 & 0.9742 &  0.5522 \\
 0.999 & 195 & 0.9818 & 0.9535 &  0.6078 \\
 0.999 & 250 & 0.9722 & 0.9240 &  0.6340 \\
\bottomhline
\end{tabular}
\end{table}

As there are two competing goals –- the inversion on one hand and J-evolution on the other –- the set goal of achieving \(q\) is actually never fully fulfilled. With higher \(q\) there are just more constrains on J-evolution as compared to the inversion properties. The quantity \(q_\mathrm{eff}\) is the amount of \(q\) that you can actually get with the pulse.

Accordingly, \(q \gtrapprox 0.6\) assigns the region which cannot bring additional J-evolution because of the opposing requirements needed for J-evolution and \(180\degree\) rotations. Since J-coupling is only acquired for transverse magnetisation, it needs to leave the transverse plane for the \(180\degree\) rotation. Accordingly, and with the additional demand for robustness at a given pulse length \(T=1~\unit{ms}\), it is concluded that the resultant J\(_\text{ev}\) pulses can acquire up to \(q \approx 0.6\) J-evolution. This corresponds to 60\% of \(T\) and allows us to reduce the duration of flanked delays by \(600~\unit{\mu s}\) (shown in Fig~.\ref{FIG_pulses}).

In addition to the J-evolution capabilities of the pulse sandwich, the performance with respect to \(180\degree\) rotations should also be compared. Whilst the inversion of \(z\)-magnetisation can be calculated easily, the full rotational performance is much more difficult to characterise. For this evaluation, the quality factor \(\Phi_\mathrm{ev}\) with \(J=0~\unit{Hz}\) may be used, in principle, because this implies vanishing \( \mathbf{R}_\mathrm{J}\). However, the rotation performance will also be J-dependent, thereby masking the real performance of the pulse sandwich when only a single coupling value is used. In the authors' experience, a better definition is obtained by considering the ensemble of performances at all couplings by the fidelity function \(\Phi_\mathrm{ens}\):
\begin{align}
& \Phi_\mathrm{ens}=\max{\Big[\mathrm{Re}\big\langle \mathbf{R}_\mathrm{ev}(J)\big| \mathbf{U}_\mathrm{eff}\big\rangle\Big]}, & J\in\big(0,250\big]. & \label{EQ_phi_ens}
\end{align}

Similar to \(\Phi_\mathrm{ev}\), \(\Phi_\mathrm{ens}\) is defined with the target propagator \(\mathbf{R}_J\) in Eq.(\ref{Eq_prop_J}) but with a difference that \(J\) is not set to \(J_\mathrm{max}\), e.g. \(J_\mathrm{max} = 195~\unit{Hz})\), but is varied in the range of \(J = \big\{0,\dots,J_\mathrm{max}\big\}\) (with a typical increment of \(\Delta J =1~\unit{Hz}\)). \(\Phi_\mathrm{ens}\) is calculated in that range for every combination of constraints with only the best transfer efficiency from the range \(J = \big\{0,\dots,J_\mathrm{max}\big\}\) taken for the accumulation of the final transfer efficiency. This procedure is equivalent to allowing every possible J-evolution and, essentially, monitoring the effect of the \(180\degree\) rotation. The resulting plot is given in Fig.~\ref{FIG_Transfer_efficiencies_1}(A-C) (dotted lines) and is shown with more detail in Fig.~\ref{FIG_Transfer_efficiencies_2} (dotted line). As a general observation, the rotational performances over all J\(_\text{ev}\) pulse sandwiches are good, with a small decrease in fidelity at larger \(q\). 

\subsection{Comparison to Pulse Sandwiches without J-Evolution}

Analysis so far has shown that it is possible to obtain J\(_\text{ev}\) pulse sandwiches that acquire J-coupling on top of a \(180\degree\) rotation. However, it is also necessary to also evaluate whether shorter, more conventional, \(180\degree\) shaped pulses, flanked by appropriate delays, are more efficient when compared to the proposed J\(_\text{ev}\) pulses.

According to Fig.~\ref{FIG_Pulse_sequences}(D), any J\(_\text{ev}\) pulse can be imagined as a concatenation of free evolution periods, \(\tau\), with a centred, concurrent \(180\degree\) pulse sandwich (which we call a \textsc{bubu}-pulse sandwich, following the nomenclature introduced in \citep{Ehni:2013}, for Broadband Universal rotation pulses on both nuclei \(^1\)H and \(^{13}\)C). Indeed, conventional shaped pulses, flanked with delays \(\tau\), can be used to give J\(_\text{ev}\)-building blocks with duration of \(1~\unit{ms}\), which can be compared with the J\(_\text{ev}\) pulse sandwich.

The most direct comparison between J\(_\text{ev}\) pulse sandwiches and corresponding J\(_\text{ev}\)-building blocks (composed of two delays \(\tau\)  surrounding a \textsc{bubu}-pulse sandwich) can be obtained by considering the \(180\degree\) rotation capabilities and using \(\Phi_\mathrm{ens}\) in Eq.~(\ref{EQ_phi_ens}). By definition, the J\(_\text{ev}\)-building block acquires the same amount of J-coupling compared to a J\(_\text{ev}\) pulse defined with a given \(q\). The analysis starts by finding appropriate delays \(\tau\) that correspond to the J-evolution acquired upon implementation of a J\(_\text{ev}\) pulse. \(\tau\) is calculated starting from a given \(q_\mathrm{eff}\). The \(q_\mathrm{eff}\) values from Fig.~\ref{FIG_Transfer_efficiencies_1}(E) (dashed line) are given in Table~\ref{tab_qeff} (for example \(q_\mathrm{eff} = 0.6078 = 2\tau\)). This results in \(T_\textsc{bubu} = T_{J_\mathrm{ev}} - 2\tau  = 1~\unit{ms} -0.6078~\unit{ms} = 0.3922~\unit{ms}\).

In total, five \textsc{bubu}-pulses are optimised for \(J = 195~\unit{Hz}\), with durations of \(T_\textsc{bubu}\in\{0.94,0.73, 0.53, 0.42, 0.39\}~\unit{ms}\) have been optimised corresponding to \(q_\mathrm{eff} = \{0.2, 0.4, 0.6, 0.8, 0.999\}\). The decreased duration of pulses allows pulses to be optimised using \(x\)-, \(y\)-controls with a digitisation of \(0.5~\unit{\mu s}\), instead of \(x\)-, \(y\)-, \(z\)-controls, further opening the available optimisation space.

\begin{figure}
\centering{\includegraphics{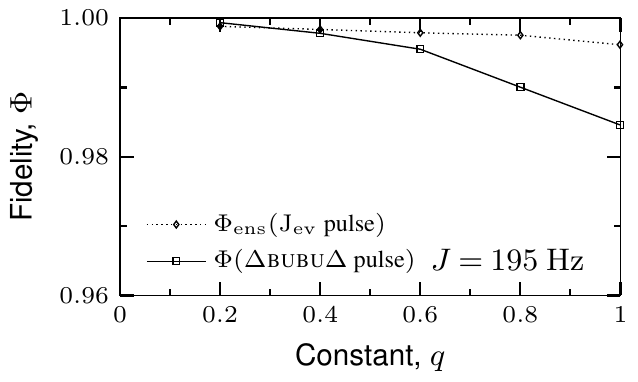}}
\caption{Transfer efficiencies for concurrent \(180\degree\) refocusing pulses, calculated for \(J = 195~\unit{Hz}\) and averaged over offsets and \(B_1\) deviations. To determine whether J\(_\text{ev}\) pulse sandwiches exceed the performance of comparable (\(\tau - 180\degree - \tau\)) J\(_\text{ev}\)-building blocks (according to the relation in Fig.~\ref{FIG_Pulse_sequences}(D)), they are compared in terms of their \(180\degree\) fidelity. Five J\(_\text{ev}\) pulses for \(0.2 \leqslant q \leqslant 0.999\) are evaluated according to \(\Phi_\mathrm{ens}\) (dotted line) and compared to J\(_\text{ev}\)-building blocks derived from shorter \textsc{bubu}-pulses. In all cases J\(_\text{ev}\) pulse sandwiches exceed the transfer properties of \textsc{bubu}-pulse sandwiches, demonstrating the gain of the new class of pulses which extend the \textsc{iceberg}-principle \citep{Gershenzon:2008}. \label{FIG_Transfer_efficiencies_2}}
\end{figure}

The fidelity of the \textsc{bubu}-pulses is obtained according to \(\Phi_\mathrm{ens}\) in Eq.~(\ref{EQ_phi_ens}) and results are shown in Fig.~\ref{FIG_Transfer_efficiencies_2} (solid line). This plot determines the physical threshold for concurrent \(180\degree\) refocusing pulses with respect to the given pulse lengths \(T_\textsc{bubu}\) and required robustness. Concerning Fig.~\ref{FIG_Transfer_efficiencies_2}, the \(180\degree\) rotation capability of the J\(_\text{ev}\) pulse sandwich (dotted line) exceeds in all cases the performance of the \textsc{bubu}-pulse sandwich (solid line). Since the setup is chosen in order to start from identical amounts of J-evolution, it is concluded that J\(_\text{ev}\) pulses are more efficient compared to an equivalent J\(_\text{ev}\)-building block made up from \textsc{bubu}-pulses and delays \(\tau\). Fig.~\ref{FIG_Transfer_efficiencies_2} shows that the efficiencies of the J\(_\text{ev}\) pulse sandwiches, compared to the \textsc{bubu} analogue, have largest gains with increasing \(q\).

\conclusions

The J\(_\text{ev}\) pulses are non-conventional pulse sandwiches that exceed the class of single spin shaped pulses. They are designed for chemical shift refocusing in heteronuclear coherence transfer elements.

Usually, delays dominate the length of typical coherence transfer elements (e.g. that of the \textsc{inept} \citep{Morris:1979} or \textsc{cob-inept} \citep{Ehni:2012}) and thereby also the total sequence duration. In order to obtain time optimal coherence transfers, it is not necessary to optimise the sequence as a whole. A fraction of the delay might be located inside the refocusing pulses, reducing the sequence length to a total duration which is closer to time-optimal.

For this reason a concept is derived from these so called J\(_\text{ev}\) pulses, with the example of J-evolved \textsc{bubu}-pulses, in that they have good refocusing properties on both nuclei, \(^1\)H and \(^{13}\)C, and also evolve J-coupling on top of these refocusing properties.

A strategy to analyse the obtained pulses is described based on the decomposition of the mutual refocusing and J-evolution properties. It is clear that the proposed J\(_\text{ev}\) pulses facilitate J-coupling evolution.

Using a second type of decomposition strategy and a set of comparable, conventional pulses, it is shown that J\(_\text{ev}\) pulses not only evolve J-coupling but they also result in sequences with reduced overall time consumption which is inaccessible with conventional pulses, and therefore closer to the physical limit of pulse sequence duration. The advantage in time consumption can be given as a fraction of the J\(_\text{ev}\) pulse length \(T = 1~\unit{ms}\) and is found to be approximately \(q_\mathrm{eff} = 0.6\) for the studied set of constraints, resulting in a reduced time demand of \(0.6~\unit{ms}\) for every J\(_\text{ev}\) pulse that is applied in a sequence, while maintaining the benefits from offset- and \(B_1\)-compensated pulse shapes.

%\dataavailability{Pulse shapes and parameters are available for download at \url{https://www.ioc.kit.edu/luy} under Downloads/Pulses.} 

%\authorcontribution{All authors contributed significantly to the work.}

%\competinginterests{None of the authors of this paper has a financial or personal relationship with other people or organisations that could inappropriately influence or bias the content of the paper.}

\begin{acknowledgements}
This project is funded by the Helmholtz programme BIFTM (47.02.04) and Information (43.35.02) and the Deutsche Forschungsgemeinschaft (DFG LU 835/11-1 and LU 835/13-1). 
\end{acknowledgements}

\bibliographystyle{copernicus}

\begin{thebibliography}{64}
\providecommand{\natexlab}[1]{#1}
\providecommand{\url}[1]{{\tt #1}}
\providecommand{\urlprefix}{URL }
\expandafter\ifx\csname urlstyle\endcsname\relax
  \providecommand{\doi}[1]{https://doi.org/\discretionary{}{}{}#1}\else
  \providecommand{\doi}{https://doi.org/\discretionary{}{}{}\begingroup
  \urlstyle{rm}\Url}\fi

\bibitem[{Bodor et~al.(2020)Bodor, Haller, Bouguechtouli, Theillet, Nyitray,
  and Luy}]{Bodor:2020}
Bodor, A., Haller, J.~D., Bouguechtouli, C., Theillet, F.-X., Nyitray, L., and
  Luy, B.: Power of Pure Shift H$\alpha$C$\alpha$ Correlations: A Way to
  Characterize Biomolecules under Physiological Conditions, Anal. Chem., 92,
  12\,423--12\,428, \doi{10.1021/acs.analchem.0c02182}, 2020.

\bibitem[{Braun and Glaser(2014)}]{Braun:2014}
Braun, M. and Glaser, S.~J.: Concurrently optimized cooperative pulses in
  robust quantum control: application to broadband Ramsey-type pulse sequence
  elements, New J. Phys., 16, 115\,002, \doi{10.1088/1367-2630/16/11/115002},
  2014.

\bibitem[{Conolly et~al.(1986)Conolly, Nishimura, and Macovski}]{Conolly:1986}
Conolly, S., Nishimura, D., and Macovski, A.: Optimal control solutions to the
  magnetic resonance selective excitation problem, {IEEE} Trans. Med. Imag., 5,
  106--115, \doi{10.1109/tmi.1986.4307754}, 1986.

\bibitem[{de~Fouquieres et~al.(2011)de~Fouquieres, Schirmer, Glaser, and
  Kuprov}]{deFouquieres:2011}
de~Fouquieres, P., Schirmer, S.~G., Glaser, S.~J., and Kuprov, I.: Second order
  gradient ascent pulse engineering, J. Magn. Reson., 212, 412--417,
  \doi{10.1016/j.jmr.2011.07.023}, 2011.

\bibitem[{Dolde et~al.(2014)Dolde, Bergholm, Wang, Jakobi, Naydenov, Pezzagna,
  Meijer, Jelezko, Neumann, Schulte-Herbr{\"{u}}ggen, Biamonte, and
  Wrachtrup}]{Dolde:2014}
Dolde, F., Bergholm, V., Wang, Y., Jakobi, I., Naydenov, B., Pezzagna, S.,
  Meijer, J., Jelezko, F., Neumann, P., Schulte-Herbr{\"{u}}ggen, T., Biamonte,
  J., and Wrachtrup, J.: High-fidelity spin entanglement using optimal control,
  Nat. Commun., 5, 3371, \doi{10.1038/ncomms4371}, 2014.

\bibitem[{Doll et~al.(2013)Doll, Pribitzer, Tschaggelar, and
  Jeschke}]{Doll:2013}
Doll, A., Pribitzer, S., Tschaggelar, R., and Jeschke, G.: Adiabatic and fast
  passage ultra-wideband inversion in pulsed {EPR}, J. Magn. Reson., 230,
  27--39, \doi{10.1016/j.jmr.2013.01.002}, 2013.

\bibitem[{Ehni and Luy(2012)}]{Ehni:2012}
Ehni, S. and Luy, B.: A systematic approach for optimizing the robustness of
  pulse sequence elements with respect to couplings, offsets, and B1-field
  inhomogeneities ({COB}), Magn. Reson. Chem., 50, 63--72,
  \doi{10.1002/mrc.3846}, 2012.

\bibitem[{Ehni and Luy(2013)}]{Ehni:2013}
Ehni, S. and Luy, B.: BEBE$^\text{tr}$ and BUBI: J-compensated concurrent
  shaped pulses for 1H--13C experiments, J. Magn. Reson., 232, 7--17,
  \doi{10.1016/j.jmr.2013.04.007}, 2013.

\bibitem[{Ehni and Luy(2014)}]{Ehni:2014}
Ehni, S. and Luy, B.: Robust {INEPT} and refocused {INEPT} transfer with
  compensation of a wide range of couplings, offsets, and B1-field
  inhomogeneities ({COB3}), J. Magn. Reson., 247, 111--117,
  \doi{10.1016/j.jmr.2014.07.010}, 2014.

\bibitem[{Emsley and Bodenhausen(1990)}]{Emsley:1990}
Emsley, L. and Bodenhausen, G.: Gaussian pulse cascades: New analytical
  functions for rectangular selective inversion and in-phase excitation in
  {NMR}, Chem. Phys. Lett., 165, 469--476, \doi{10.1016/0009-2614(90)87025-m},
  1990.

\bibitem[{Ewing et~al.(1990)Ewing, Glaser, and Drobny}]{Ewing:1990}
Ewing, B., Glaser, S.~J., and Drobny, G.~P.: Development and optimization of
  shaped {NMR} pulses for the study of coupled spin systems, J. Chem. Phys.,
  147, 121--129, \doi{10.1016/0301-0104(90)85028-u}, 1990.

\bibitem[{Freeman et~al.(1988)Freeman, Friedrich, and Xi-li}]{Freeman:1988}
Freeman, R., Friedrich, J., and Xi-li, W.: A pulse for all seasons. Fourier
  transform spectra without a phase gradient, J. Magn. Reson. (1969), 79,
  561--567, \doi{10.1016/0022-2364(88)90092-3}, 1988.

\bibitem[{Garwood and Ke(1991)}]{Garwood:1991}
Garwood, M. and Ke, Y.: Symmetric pulses to induce arbitrary flip angles with
  compensation for rf inhomogeneity and resonance offsets, J. Magn. Reson.
  (1969), 94, 511--525, \doi{10.1016/0022-2364(91)90137-i}, 1991.

\bibitem[{Gershenzon et~al.(2008)Gershenzon, Skinner, Brutscher, Khaneja,
  Nimbalkar, Luy, and Glaser}]{Gershenzon:2008}
Gershenzon, N.~I., Skinner, T.~E., Brutscher, B., Khaneja, N., Nimbalkar, M.,
  Luy, B., and Glaser, S.~J.: Linear phase slope in pulse design: Application
  to coherence transfer, J. Magn. Reson., 192, 235--243,
  \doi{10.1016/j.jmr.2008.02.021}, 2008.

\bibitem[{Glaser et~al.(2015)Glaser, Boscain, Calarco, Koch,
  K{\"{o}}ckenberger, Kosloff, Kuprov, Luy, Schirmer, Schulte-Herbr{\"{u}}ggen,
  Sugny, and Wilhelm}]{Glaser:2015}
Glaser, S.~J., Boscain, U., Calarco, T., Koch, C.~P., K{\"{o}}ckenberger, W.,
  Kosloff, R., Kuprov, I., Luy, B., Schirmer, S., Schulte-Herbr{\"{u}}ggen, T.,
  Sugny, D., and Wilhelm, F.~K.: Training Schr{\"{o}}dinger's cat: quantum
  optimal control, Eur. Phys. J. D, 69, 279, \doi{10.1140/epjd/e2015-60464-1},
  2015.

\bibitem[{Goodwin and Kuprov(2015)}]{Goodwin:2015}
Goodwin, D.~L. and Kuprov, I.: Auxiliary matrix formalism for interaction
  representation transformations, optimal control, and spin relaxation
  theories, J. Chem. Phys., 143, 084\,113, \doi{10.1063/1.4928978}, 2015.

\bibitem[{Goodwin and Kuprov(2016)}]{Goodwin:2016}
Goodwin, D.~L. and Kuprov, I.: Modified Newton-Raphson {GRAPE} methods for
  optimal control of spin systems, J. Chem. Phys., 144, 204\,107,
  \doi{10.1063/1.4949534}, 2016.

\bibitem[{Goodwin et~al.(2018)Goodwin, Myers, Timmel, and
  Kuprov}]{Goodwin:2018}
Goodwin, D.~L., Myers, W.~K., Timmel, C.~R., and Kuprov, I.: Feedback control
  optimisation of ESR experiments, J. Magn. Reson., 297, 9--16,
  \doi{10.1016/j.jmr.2018.09.009}, 2018.

\bibitem[{Goodwin et~al.(2020)Goodwin, Koos, and Luy}]{Goodwin:2020}
Goodwin, D.~L., Koos, M. R.~M., and Luy, B.: Second Order Phase Dispersion by
  Optimised Rotations, Phys. Rev. Res., 2, 033\,157,
  \doi{10.1103/PhysRevResearch.2.033157}, 2020.

\bibitem[{Haller et~al.(2019)Haller, Bodor, and Luy}]{Haller:2019}
Haller, J.~D., Bodor, A., and Luy, B.: Real-time pure shift measurements for
  uniformly isotope-labeled molecules using {X}-selective {BIRD} homonuclear
  decoupling, J. Magn. Reson., 302, 64--71, \doi{10.1016/j.jmr.2019.03.011},
  2019.

\bibitem[{Hohenester and Stadler(2005)}]{Hohenester:2005}
Hohenester, U. and Stadler, G.: Optimal quantum control of electron-phonon
  scatterings in artificial atoms, Physica E, 29, 320,
  \doi{10.1016/j.physe.2005.05.029}, 2005.

\bibitem[{Janich et~al.(2011)Janich, Schulte, Schwaiger, and
  Glaser}]{Janich:2011}
Janich, M.~A., Schulte, R.~F., Schwaiger, M., and Glaser, S.~J.: Robust
  slice-selective broadband refocusing pulses, J. Magn. Reson., 213, 126--135,
  \doi{10.1016/j.jmr.2011.09.025}, 2011.

\bibitem[{Kaufmann et~al.(2013)Kaufmann, Keller, Franck, Barnes, Glaser,
  Martinis, and Han}]{Kaufmann:2013}
Kaufmann, T., Keller, T.~J., Franck, J.~M., Barnes, R.~P., Glaser, S.~J.,
  Martinis, J.~M., and Han, S.: {DAC}-board based {X}-band {EPR} spectrometer
  with arbitrary waveform control, J. Magn. Reson., 235, 95--108,
  \doi{10.1016/j.jmr.2013.07.015}, 2013.

\bibitem[{Khaneja and Glaser(2002)}]{Khaneja:2002b}
Khaneja, N. and Glaser, S.~J.: Efficient transfer of coherence through Ising
  spin chains, Phys. Rev. A, 66, 060\,301, \doi{10.1103/physreva.66.060301},
  2002.

\bibitem[{Khaneja et~al.(2001)Khaneja, Brockett, and Glaser}]{Khaneja:2001}
Khaneja, N., Brockett, R., and Glaser, S.~J.: Time optimal control in spin
  systems, Phys. Rev. A, 63, 032\,308, \doi{10.1103/physreva.63.032308}, 2001.

\bibitem[{Khaneja et~al.(2002)Khaneja, Glaser, and Brockett}]{Khaneja:2002a}
Khaneja, N., Glaser, S.~J., and Brockett, R.: Sub-Riemannian geometry and time
  optimal control of three spin systems: Quantum gates and coherence transfer,
  Phys. Rev. A, 65, \doi{10.1103/physreva.65.032301}, 2002.

\bibitem[{Khaneja et~al.(2003{\natexlab{a}})Khaneja, Luy, and
  Glaser}]{Khaneja:2003b}
Khaneja, N., Luy, B., and Glaser, S.~J.: Boundary of quantum evolution under
  decoherence, Proc. Natl. Acad. Sci., 100, 13\,162--13\,166,
  \doi{10.1073/pnas.2134111100}, 2003{\natexlab{a}}.

\bibitem[{Khaneja et~al.(2003{\natexlab{b}})Khaneja, Reiss, Luy, and
  Glaser}]{Khaneja:2003a}
Khaneja, N., Reiss, T., Luy, B., and Glaser, S.~J.: Optimal control of spin
  dynamics in the presence of relaxation, J. Magn. Reson., 162, 311--319,
  \doi{10.1016/s1090-7807(03)00003-x}, 2003{\natexlab{b}}.

\bibitem[{Khaneja et~al.(2005)Khaneja, Reiss, Kehlet, Schulte-Herbr{\"{u}}ggen,
  and Glaser}]{Khaneja:2005}
Khaneja, N., Reiss, T., Kehlet, C., Schulte-Herbr{\"{u}}ggen, T., and Glaser,
  S.~J.: Optimal control of coupled spin dynamics: design of {NMR} pulse
  sequences by gradient ascent algorithms, J. Magn. Reson., 172, 296--305,
  \doi{10.1016/j.jmr.2004.11.004}, 2005.

\bibitem[{Kobzar(2007)}]{Kobzar_thesis}
Kobzar, K.: Optimal Control, Partial Alignment and More: The Design Of Novel
  Tools for {NMR} Spectroscopy of Small Molecules, Ph.D. thesis, Technische
  Universit{\"a}t M{\"u}nchen, 2007.

\bibitem[{Kobzar et~al.(2004)Kobzar, Skinner, Khaneja, Glaser, and
  Luy}]{Kobzar:2004}
Kobzar, K., Skinner, T.~E., Khaneja, N., Glaser, S.~J., and Luy, B.: Exploring
  the limits of broadband excitation and inversion pulses, J. Magn. Reson.,
  170, 236--243, \doi{10.1016/j.jmr.2004.06.017}, 2004.

\bibitem[{Kobzar et~al.(2005)Kobzar, Luy, Khaneja, and Glaser}]{Kobzar:2005}
Kobzar, K., Luy, B., Khaneja, N., and Glaser, S.~J.: Pattern pulses: design of
  arbitrary excitation profiles as a function of pulse amplitude and offset, J.
  Magn. Reson., 173, 229--235, \doi{10.1016/j.jmr.2004.12.005}, 2005.

\bibitem[{Kobzar et~al.(2008)Kobzar, Skinner, Khaneja, Glaser, and
  Luy}]{Kobzar:2008}
Kobzar, K., Skinner, T.~E., Khaneja, N., Glaser, S.~J., and Luy, B.: Exploring
  the limits of broadband excitation and inversion: {II}. Rf-power optimized
  pulses, J. Magn. Reson., 194, 58--66, \doi{10.1016/j.jmr.2008.05.023}, 2008.

\bibitem[{Kobzar et~al.(2012)Kobzar, Ehni, Skinner, Glaser, and
  Luy}]{Kobzar:2012}
Kobzar, K., Ehni, S., Skinner, T.~E., Glaser, S.~J., and Luy, B.: Exploring the
  limits of broadband 90$^{\circ}$ and 180$^{\circ}$ universal rotation pulses,
  J. Magn. Reson., 225, 142--160, \doi{10.1016/j.jmr.2012.09.013}, 2012.

\bibitem[{Koos et~al.(2015)Koos, Feyrer, and Luy}]{Koos:2015}
Koos, M. R.~M., Feyrer, H., and Luy, B.: Broadband excitation pulses with
  variable {RF} amplitude-dependent flip angle ({RADFA}), Magn. Reson. Chem.,
  53, 886--893, \doi{10.1002/mrc.4297}, 2015.

\bibitem[{Koos et~al.(2017)Koos, Feyrer, and Luy}]{Koos:2017}
Koos, M. R.~M., Feyrer, H., and Luy, B.: Broadband {RF}-amplitude-dependent
  flip angle pulses with linear phase slope, Magn. Reson. Chem., 55, 797--803,
  \doi{10.1002/mrc.4593}, 2017.

\bibitem[{Lescop et~al.(2010)Lescop, Kern, and Brutscher}]{Lescop:2010}
Lescop, E., Kern, T., and Brutscher, B.: Guidelines for the use of
  band-selective radiofrequency pulses in hetero-nuclear {NMR}: example of
  longitudinal-relaxation-enhanced {BEST}-type {1H}--{15N} correlation
  experiments, J. Magn. Reson., 203, 190--198, \doi{10.1016/j.jmr.2009.12.001},
  2010.

\bibitem[{Levante et~al.(1996)Levante, Bremi, and Ernst}]{Levante:1996}
Levante, T.~O., Bremi, T., and Ernst, R.~R.: Pulse-sequence optimization with
  analytical derivatives. Application to deuterium decoupling in oriented
  phases, J. Magn. Reson., A, 121, 167--177, \doi{10.1006/jmra.1996.0157},
  1996.

\bibitem[{Levitt(1982)}]{Levitt:1982}
Levitt, M.~H.: Symmetrical composite pulse sequences for {NMR} population
  inversion. I. Compensation of radiofrequency field inhomogeneity, J. Magn.
  Reson. (1969), 48, 234--264, \doi{10.1016/0022-2364(82)90275-x}, 1982.

\bibitem[{Levitt(1986)}]{Levitt:1986}
Levitt, M.~H.: Composite Pulses, Prog. Nucl. Magn. Reson. Spectrosc., 18,
  61--122, \doi{10.1016/0079-6565(86)80005-X}, 1986.

\bibitem[{Lingel et~al.(2020)Lingel, Vulpetti, Reinsperger, Proudfoot, Denay,
  Frommlet, Henry, Hommel, Gossert, Luy, and Frank}]{Lingel:2020}
Lingel, A., Vulpetti, A., Reinsperger, T., Proudfoot, A., Denay, R., Frommlet,
  A., Henry, C., Hommel, U., Gossert, A.~D., Luy, B., and Frank, A.~O.:
  Comprehensive and High-Throughput Exploration of Chemical Space Using
  Broadband {19F} {NMR}-Based Screening, Angew. Chem., 59, 14\,809--14\,817,
  \doi{10.1002/anie.202002463}, 2020.

\bibitem[{Lurie(1985)}]{Lurie:1985}
Lurie, D.~J.: A systematic design procedure for selective pulses in {NMR}
  imaging, Magn. Reson. Imag., 3, 235--243, \doi{10.1016/0730-725x(85)90352-2},
  1985.

\bibitem[{Machnes et~al.(2018)Machnes, Ass{\'{e}}mat, Tannor, and
  Wilhelm}]{Machnes:2018}
Machnes, S., Ass{\'{e}}mat, E., Tannor, D., and Wilhelm, F.~K.: Tunable,
  Flexible, and Efficient Optimization of Control Pulses for Practical Qubits,
  Phys. Rev. Lett., 120, 150\,401, \doi{10.1103/physrevlett.120.150401}, 2018.

\bibitem[{Morris and Freeman(1979)}]{Morris:1979}
Morris, G.~A. and Freeman, R.: Enhancement of nuclear magnetic resonance
  signals by polarization transfer, J. Am. Chem. Soc., 101, 760,
  \doi{10.1021/ja00497a058}, 1979.

\bibitem[{Ndong and Koch(2010)}]{Ndong:2010}
Ndong, M. and Koch, C.~P.: Vibrational stabilization of ultracold KRb
  molecules: A comparative study, Phys. Rev. A, 83, 43\,437,
  \doi{10.1103/PhysRevA.82.043437}, 2010.

\bibitem[{Palao et~al.(2008)Palao, Kosloff, and Koch}]{Palao:2008}
Palao, J.~P., Kosloff, R., and Koch, C.~P.: Protecting coherence in optimal
  control theory: State-dependent constraint approach, Phys. Rev. A, 77,
  63\,412, \doi{10.1103/PhysRevA.77.063412}, 2008.

\bibitem[{Reiss et~al.(2002)Reiss, Khaneja, and Glaser}]{Reiss:2002}
Reiss, T.~O., Khaneja, N., and Glaser, S.~J.: Time-Optimal
  Coherence-Order-Selective Transfer of In-Phase Coherence in Heteronuclear
  {IS} Spin Systems, J. Magn. Reson., 154, 192--195,
  \doi{10.1006/jmre.2001.2480}, 2002.

\bibitem[{Reiss et~al.(2003)Reiss, Khaneja, and Glaser}]{Reiss:2003}
Reiss, T.~O., Khaneja, N., and Glaser, S.~J.: Broadband geodesic pulses for
  three spin systems: time-optimal realization of effective trilinear coupling
  terms and indirect {SWAP} gates, J. Magn. Reson., 165, 95--101,
  \doi{10.1016/s1090-7807(03)00245-3}, 2003.

\bibitem[{Saywell et~al.(2018)Saywell, Kuprov, Goodwin, Carey, and
  Freegarde}]{Saywell:2018}
Saywell, J.~C., Kuprov, I., Goodwin, D., Carey, M., and Freegarde, T.: Optimal
  control of mirror pulses for cold-atom interferometry, Phys. Rev. A, 98,
  023\,625, \doi{10.1103/physreva.98.023625}, 2018.

\bibitem[{Shaka and Freeman(1983)}]{Shaka:1983}
Shaka, A. and Freeman, R.: Composite pulses with dual compensation, J. Magn.
  Reson. (1969), 55, 487--493, \doi{10.1016/0022-2364(83)90133-6}, 1983.

\bibitem[{Shaka and Pines(1987)}]{Shaka:1987}
Shaka, A. and Pines, A.: Symmetric phase-alternating composite pulses, J. Magn.
  Reson. (1969), 71, 495--503, \doi{10.1016/0022-2364(87)90249-6}, 1987.

\bibitem[{Skinner et~al.(2003)Skinner, Reiss, Luy, Khaneja, and
  Glaser}]{Skinner:2003}
Skinner, T.~E., Reiss, T.~O., Luy, B., Khaneja, N., and Glaser, S.~J.:
  Application of optimal control theory to the design of broadband excitation
  pulses for high-resolution {NMR}, J. Magn. Reson., 163, 8--15,
  \doi{10.1016/S1090-7807(03)00153-8}, 2003.

\bibitem[{Skinner et~al.(2004)Skinner, Reiss, Luy, Khaneja, and
  Glaser}]{Skinner:2004}
Skinner, T.~E., Reiss, T.~O., Luy, B., Khaneja, N., and Glaser, S.~J.: Reducing
  the duration of broadband excitation pulses using optimal control with
  limited {RF} amplitude, J. Magn. Reson., 167, 68--74,
  \doi{10.1016/j.jmr.2003.12.001}, 2004.

\bibitem[{Skinner et~al.(2005)Skinner, Reiss, Luy, Khaneja, and
  Glaser}]{Skinner:2005}
Skinner, T.~E., Reiss, T.~O., Luy, B., Khaneja, N., and Glaser, S.~J.:
  Tailoring the optimal control cost function to a desired output: application
  to minimizing phase errors in short broadband excitation pulses, J. Magn.
  Reson., 172, 17--23, \doi{10.1016/j.jmr.2004.09.011}, 2005.

\bibitem[{Skinner et~al.(2006)Skinner, Kobzar, Luy, Bendall, Bermel, Khaneja,
  and Glaser}]{Skinner:2006}
Skinner, T.~E., Kobzar, K., Luy, B., Bendall, M.~R., Bermel, W., Khaneja, N.,
  and Glaser, S.~J.: Optimal control design of constant amplitude
  phase-modulated pulses: Application to calibration-free broadband excitation,
  J. Magn. Reson., 179, 241--249, \doi{10.1016/j.jmr.2005.12.010}, 2006.

\bibitem[{Skinner et~al.(2012)Skinner, Gershenzon, Nimbalkar, Bermel, Luy, and
  Glaser}]{Skinner:2012}
Skinner, T.~E., Gershenzon, N.~I., Nimbalkar, M., Bermel, W., Luy, B., and
  Glaser, S.~J.: New strategies for designing robust universal rotation pulses:
  Application to broadband refocusing at low power, J. Magn. Reson., 216,
  78--87, \doi{10.1016/j.jmr.2012.01.005}, 2012.

\bibitem[{Spindler et~al.(2012)Spindler, Zhang, Endeward, Gershernzon, Skinner,
  Glaser, and Prisner}]{Spindler:2012}
Spindler, P.~E., Zhang, Y., Endeward, B., Gershernzon, N., Skinner, T.~E.,
  Glaser, S.~J., and Prisner, T.~F.: Shaped optimal control pulses for
  increased excitation bandwidth in {EPR}, J. Magn. Reson., 218, 49--58,
  \doi{10.1016/j.jmr.2012.02.013}, 2012.

\bibitem[{Stefanatos et~al.(2004)Stefanatos, Khaneja, and
  Glaser}]{Stefanatos:2004}
Stefanatos, D., Khaneja, N., and Glaser, S.~J.: Optimal control of coupled
  spins in the presence of longitudinal and transverse relaxation, Phys. Rev.
  A, 69, 022\,319, \doi{10.1103/physreva.69.022319}, 2004.

\bibitem[{To{\v{s}}ner et~al.(2006)To{\v{s}}ner, Glaser, Khaneja, and
  Nielsen}]{Tosner:2006}
To{\v{s}}ner, Z., Glaser, S.~J., Khaneja, N., and Nielsen, N.~C.: Effective
  Hamiltonians by optimal control: Solid-state {NMR} double-quantum planar and
  isotropic dipolar recoupling, J. Chem. Phys., 125, 184\,502,
  \doi{10.1063/1.2366703}, 2006.

\bibitem[{Tycko et~al.(1985)Tycko, Cho, Schneider, and Pines}]{Tycko:1985}
Tycko, R., Cho, H., Schneider, E., and Pines, A.: Composite pulses without
  phase distortion, J. Magn. Reson. (1969), 61, 90--101,
  \doi{10.1016/0022-2364(85)90270-7}, 1985.

\bibitem[{{Van Reeth} et~al.(2017){Van Reeth}, Ratiney, Tesch, Grenier, Beuf,
  Glaser, and Sugny}]{VanReeth:2017}
{Van Reeth}, E., Ratiney, H., Tesch, M., Grenier, D., Beuf, O., Glaser, S.~J.,
  and Sugny, D.: Optimal control design of preparation pulses for contrast
  optimization in {MRI}, J. Magn. Reson., 279, 39--50,
  \doi{10.1016/j.jmr.2017.04.012}, 2017.

\bibitem[{Vinding et~al.(2012)Vinding, Maximov, To{\v{s}}ner, and
  Nielsen}]{Vinding:2012}
Vinding, M.~S., Maximov, I.~I., To{\v{s}}ner, Z., and Nielsen, N.~C.: Fast
  numerical design of spatial-selective rf pulses in {MRI} using Krotov and
  quasi-Newton based optimal control methods, J. Chem. Phys., 137, 054\,203,
  \doi{10.1063/1.4739755}, 2012.

\bibitem[{Vinding et~al.(2017)Vinding, Brenner, Tse, Vellmer, Vosegaard, Suter,
  St\"{o}cker, and Maximov}]{Vinding:2017}
Vinding, M.~S., Brenner, D., Tse, D. H.~Y., Vellmer, S., Vosegaard, T., Suter,
  D., St\"{o}cker, T., and Maximov, I.~I.: Application of the limited-memory
  quasi-Newton algorithm for multi-dimensional, large flip-angle {RF} pulses at
  7T, Magn. Reson. Mater. Phy., 30, 29--39, \doi{10.1007/s10334-016-0580-1},
  2017.

\bibitem[{Warren(1984)}]{Warren:1984}
Warren, W.~S.: Effects of arbitrary laser or {NMR} pulse shapes on population
  inversion and coherence, J. Chem. Phys., 81, 5437--5448,
  \doi{10.1063/1.447644}, 1984.

\end{thebibliography}

\end{document}